\documentclass[aps,prl,twocolumn,groupedaddress]{revtex4-1}
\usepackage{graphicx}
\usepackage{textcomp}
\usepackage{gensymb}
\usepackage{amsmath}
\usepackage{natbib}
\usepackage{hyperref}

\begin{document}
	
	\title{An optomechanical platform with a 3-dimensional waveguide cavity}
	
	\author{Bindu Gunupudi}
	\email[]{bindu@iisc.ac.in}
	\author{Soumya Ranjan Das}
	\author{Rohit Navarathna}
	\author{Sudhir Kumar Sahu}
	\author{Sourav Majumder}
	\author{Vibhor Singh}
	\email[]{v.singh@iisc.ac.in}	
	\affiliation{Department of Physics, Indian Institute of Science, Bangalore 560012 (India)}
	
	\date{\today}
	
	\begin{abstract}

At low temperatures, microwave cavities are often preferred for the readout and control of a variety of systems. In this paper, we present design and measurements on an optomechanical device based on a 3-dimensional rectangular waveguide cavity. We show that by suitably modifying the electromagnetic field corresponding to the fundamental mode of the cavity, the equivalent circuit capacitance can be reduced to 29~fF. By coupling a mechanical resonator to the modified electromagnetic mode of the cavity, we achieved a capacitance participation ratio of 43~$\%$. We demonstrate an optomechanical cooperativity, $C$~$\sim$~40, characterized by performing measurements in the optomechanically-induced absorption (OMIA) limit. In addition, due to a low-impedance environment between the two-halves of the cavity, our design has the flexibility of incorporating a DC bias across the mechanical resonator, often a desired feature in tunable optomechanical devices.
	\end{abstract}
	
	%\keywords{}
	
	\maketitle

\section{Introduction}  \label{section:Introduction}  

Cavity-optomechanical systems have demonstrated an exquisite ability to control the quantum state of massive-mechanical resonators.
A prototypical cavity-optomechanical system consists of an electromagnetic mode and a mechanical mode coupled by radiation-pressure or coulombic-force. \cite{braginsky_quantum_1995,RevModPhys.86.1391}. Apart from improving the transduction sensitivity \cite{ metcalfe2014applications}, these systems offer a platform to control and manipulate the quantum state of macroscopic mechanical resonators, and to implement tailored interactions \cite{brendel_pseudomagnetic_2017}. From earlier demonstrations of achieving quantum ground state, both in optical and in microwave domains \cite{teufel_sideband_2011,chan_laser_2011}, design and technological advancements in these experimental systems have enabled several milestones such as, capturing a single-microwave photon and its storage in mechanical vibrations \cite{reed_faithful_2017}, non-reciprocal microwave circuits \cite{peterson_demonstration_2017,bernier_nonreciprocal_2017}, hybrid systems with artificial atoms \cite{pirkkalainen_hybrid_2013,schmidt_ultrawide-range_2018}, and quantum entanglement between two mechanical resonators \cite{ockeloen-korppi_stabilized_2018,riedinger_remote_2018}.

For optomechanical systems in microwave domain, a common strategy is to couple a mechanically-compliant capacitor to a lumped inductor, or to the distributed inductance of a coplanar waveguide cavity \cite{teufel_circuit_2011,suh_mechanically_2014,andrews_bidirectional_2014,teufel_nanomechanical_2009,singh2014optomechanical}. In this context, 3-dimensional (3D) waveguide cavities provide a unique platform. Due to their intrinsically higher coherence, there has been a lot of interest in using 3D cavities for quantum devices \cite{reagor_reaching_2013}. Such cavities have been used for the readout and manipulation of a wide variety of systems such as superconducting qubits \cite{paik_observation_2011}, hybrid systems for inter-frequency convertors \cite{menke_reconfigurable_2017}, nitrogen-vacancy centers in diamond \cite{ball_loop-gap_2018}, and magnons \cite{tabuchi_coherent_2015}. To achieve efficient coupling of these systems to electromagnetic field, often one has to engineer the electromagnetic mode to balance the electric and magnetic field components. Primarily due to a large mode volume, 3D cavities have also been used to couple the motion of mm-sized SiN-membrane resonators demonstrating high cooperativity and quantum ground state of motion \cite{yuan2015large,noguchi2016ground}. However, in these systems, large parasitic capacitance in the microwave design, and non-lithographic device assembly techniques limit the achieved single photon coupling strength.

In this work, we demonstrate an optomechanical platform based on a 3D-waveguide cavity, where the resonant mode can be engineered to minimize the parasitic capacitance, and thereby achieve a high participation of the mechanically-compliant capacitance. These design guidelines can further be used to develop similar systems with novel materials such as graphene, NbSe$_2$, and to develop hybrid systems with other superconducting circuits while reducing fabrication complexities of the microwave part \cite{singh2014optomechanical,will_high_2017,northeast_suspension_2018}. Moreover, our optomechanical system allows to incorporate a DC voltage bias across the mechanically compliant capacitor, thus providing a tunability of mechanical and cavity modes without requiring additional on-chip filter circuitry. The design and low temperature measurements on such a system are described in the following sections.
	
\section{Simulations and Design} \label{section:Design}

\begin{figure*}
	\includegraphics[width = 170 mm]{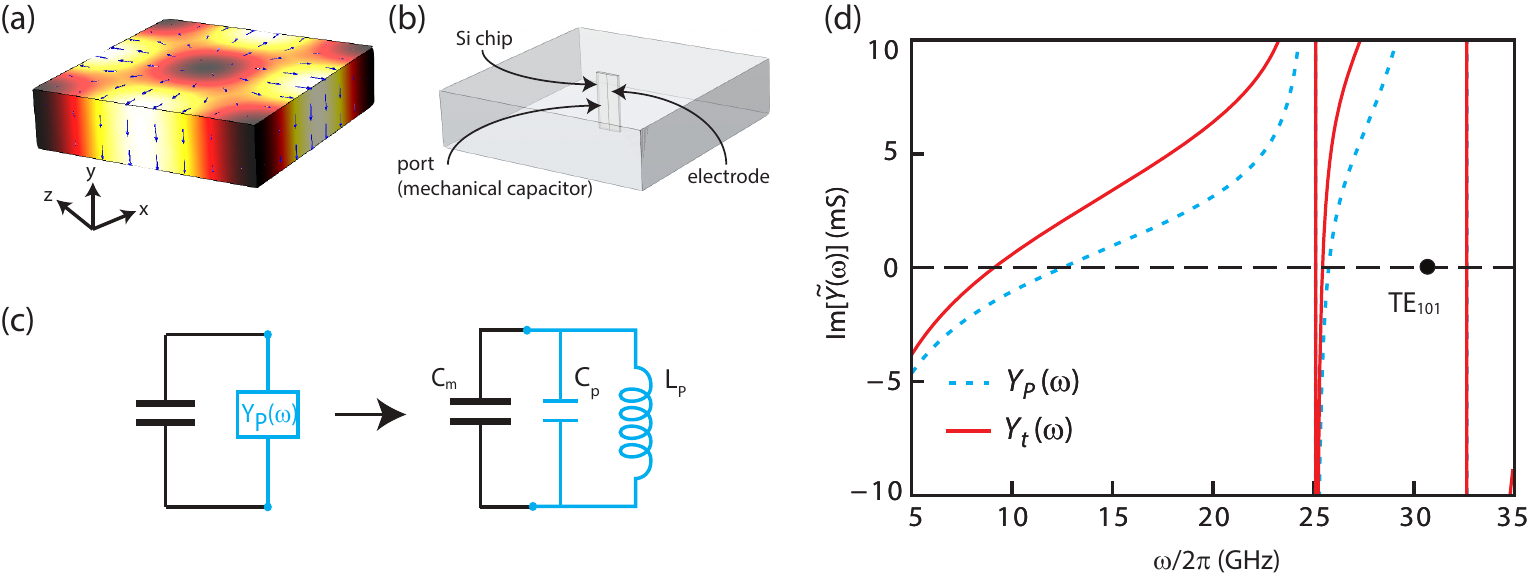}
	\caption{\textbf{Simulation and design}: (Color online)(a) An intensity colormap plot of the magnitude of charge current flowing on the surface of a rectangular waveguide cavity for the fundamental mode. White (black) color represents the maximum (minimum) current. The arrows overlaid on the surface show the direction of charge-current at some instance of time. (b) A schematic representation of simulation model, showing a silicon chip placed inside the cavity. The chip contains two electrodes running from the center of the top and bottom surfaces of the cavity and a lumped port at the centre. (c) The equivalent model for environmental admittance $\tilde{Y}_p(\omega)$ seen by a port placed between the two connecting electrodes. The additional capacitance $C_m$ represents contribution from the mechanical resonator. On the right-side: decomposition of this lumped-distributive mode into equivalent circuit elements, $C_p$ and $L_p$. (d) Plots of the imaginary part of admittance as seen by the mechanical capacitor and total admittance defined as $Y_t(\omega) = Y_p(\omega) + \omega C_m$. Each zero-crossing with a positive slope corresponds to a resonant mode.}\label{fig-1}
\end{figure*}

The central idea behind this implementation is to modify the electromagnetic field corresponding to the fundamental TE$_{101}$ mode of a rectangular waveguide cavity in a manner that maximizes the electric field at its center. A typical distribution of surface charge-current for the fundamental mode is shown in Fig.~\ref{fig-1}(a). The magnitude of the current drops towards the center of the top and bottom surfaces of the cavity. This is balanced by a corresponding displacement current flowing across the bottom and top surfaces. This suggests that one possible way to maximize the electric field at the center of the cavity could be to direct the charge current towards this region. This is achieved by electrically connecting two electrodes from the middle of the two surfaces (top and bottom), and bringing them closer towards the center, as shown schematically in Fig.~\ref{fig-1}(b). Thus, an optomechanical system can be formed by placing a mechanically compliant capacitor at the center of the cavity and galvanically connecting its leads to the cavity walls.

Fig.~\ref{fig-1}(c) shows the equivalent 
lumped element model, where the ``modified-cavity'' 
environment is represented as a complex admittance. The added 
mechanically compliant capacitor is simulated as a capacitance $C_{m}$, which further lowers
the resonant frequency of the mode. By applying Foster's theorem for a loss-less network, the complex-admittance offered by the cavity environment can be decomposed to the equivalent 
lumped elements \cite{foster1924reactance}. The mode frequencies $\omega_0$'s of the system are 
determined by the zeros of the imaginary part of the total admittance, defined by 
$Y_t(\omega) = \text{Im}[\tilde{Y}_t(\omega)] =  Y_p(\omega) + \omega C_m $\cite{foster1924reactance,nigg_black-box_2012}.
Further, the mode impedance is determined from the positive slope at zero crossing, 
i.e., $Z = \frac{2}{\omega_0}\left(\frac{d}{d\omega} (\text{Im}[\tilde{Y}_{t}(\omega)])|_{\omega_0}\right)^{-1}$. 
The capacitance $C_{p}$ and inductance $L_{p}$ of the corresponding mode are then calculated from $Z$ and $\omega_{0}$.

We perform numerical simulations to determine the eigen-frequencies 
with such a major modification to the cavity geometry
that extends beyond perturbative analysis \cite{nigg_black-box_2012}. 
With an objective to minimize the parasitic capacitance (cavity-mode capacitance $C_p$), we start with 
a 7~mm $\times$ 4~mm $\times $ 7~mm (dimensions along the x, y, z 
axes as shown in Fig.~\ref{fig-1}(a)) cavity which has a fundamental resonance mode at 30.3~GHz. We then incorporate a 3~mm~$\times$~4~mm~$\times$~0.3~mm (corresponding to the x, y, z axes) silicon chip at the center of the cavity. 
The chip is designed with a 7~$\mu$m wide electrode having a 25~$\mu$m gap at the middle of the cavity. The electrode runs 
all the way up to the cavity surfaces, forming an electrical short.
In the gap between the electrode pattern, we define a lumped port to simulate the admittance seen by the mechanical resonator. The complete geometry is represented in Fig.~\ref{fig-1}(b) and an image of a real system is shown in Fig.~\ref{fig-2}(a).  

Using finite element methods, we then numerically compute the complex admittance of the environment
$\tilde{Y}_p(\omega)$ as seen from this port. In a real device, of course, the port is 
replaced by a mechanical capacitor and an effective electromagnetic cavity mode is formed 
by combining a mechanical capacitance $C_m$ to the complex admittance $\tilde{Y}_p(\omega)$.
Since the typical size of the mechanical capacitor is much smaller than the 
wavelength of the microwave signals, it justifies the choice of a lumped port at the center.

\begin{table*}[htbp]
	\begin{center}
		\begin{tabular}{| c | c | c | c | c | c |  }
			\hline
			Cavity size  & Electrode width  & $C_{m}$ & Mode frequency & $C_{p}$   & $L_{p}$  \\ 
			(mm$^{3}$) & ($\mu$m) & (fF) &  (GHz) & (fF) & (nH)\\ \hline
			28 $\times$ 6 $\times$ 28 (for device in Fig.~\ref{fig-4})  & 50 & 7.1 & 6.4  & 148.6  & 4.0  \\ \hline
			28 $\times$ 6 $\times$ 28 & 7  & 7.1  & 6.3  & 93.7  & 6.3  \\  \hline
			7 $\times$ 4 $\times$ 7 (for device 1 in Fig.~\ref{fig-3}(c))  & 7  & 23 & 9.4  & 29.8 & 5.4 \\  
			\hline
		\end{tabular}
	\end{center}
	\caption{Summary of simulation results for different cavities and variation in the connecting electrode width. The resonant frequency of the mode after adding the mechanical capacitance $C_{m}$ is also listed. Smaller cavities with thinner connecting electrodes offer lower parasitic capacitance $C_p$. The listed values of $C_m$ are the expected mechanical capacitance from the devices studied here.}
	\label{Table1}
\end{table*}

The simulations were performed for various cavity dimensions and device geometries, in order to have a system with a maximum participation ratio. Fig.~\ref{fig-1}(d) shows plots of the imaginary parts of admittance of the cavity environment $Y_p(\omega)$, and total admittance $Y_t(\omega)$ in presence of an added mechanical capacitance of 26~fF for the dimensions mentioned earlier. The connecting electrodes have an effect of lowering 
the bare cavity TE$_{101}$ mode to 12.5~GHz. The frequency gets reduced 
further to 9.1~GHz when the mechanical capacitance is added, as indicated 
by the first zero-crossing with a positive slope in Fig.~\ref{fig-1}(d). From the 
slope at the first root of $Y_{t}(\omega)$, we estimated a cavity capacitance 
$C_p \sim 29$~fF and an inductance $L_p \sim 5.43$~nH. For an optomechanical device with $C_{m} = 23$~fF, it corresponds
to a mode impedance of 485~$\Omega$. A summary of simulation results for cavities of different sizes and electrode widths is listed in Table~1. Additional simulation results are included in the supplemental material (SM)\cite{SM}.

It is interesting to point out here that a large lumped capacitance 
at the center of the cavity leads to a significant modification of 
the eigen-modes such that most of the electric field remains concentrated 
between the lumped capacitor plates. However, the mode still retains some 
characteristics of a waveguide field distribution, and can be called a 
lumped-distributive mode, corroborating well with 
higher harmonics appearing with large spectral range as seen in Fig.~\ref{fig-1}(d).

\section{Device fabrication and assembly}

\begin{figure}
	\includegraphics[width = 85 mm]{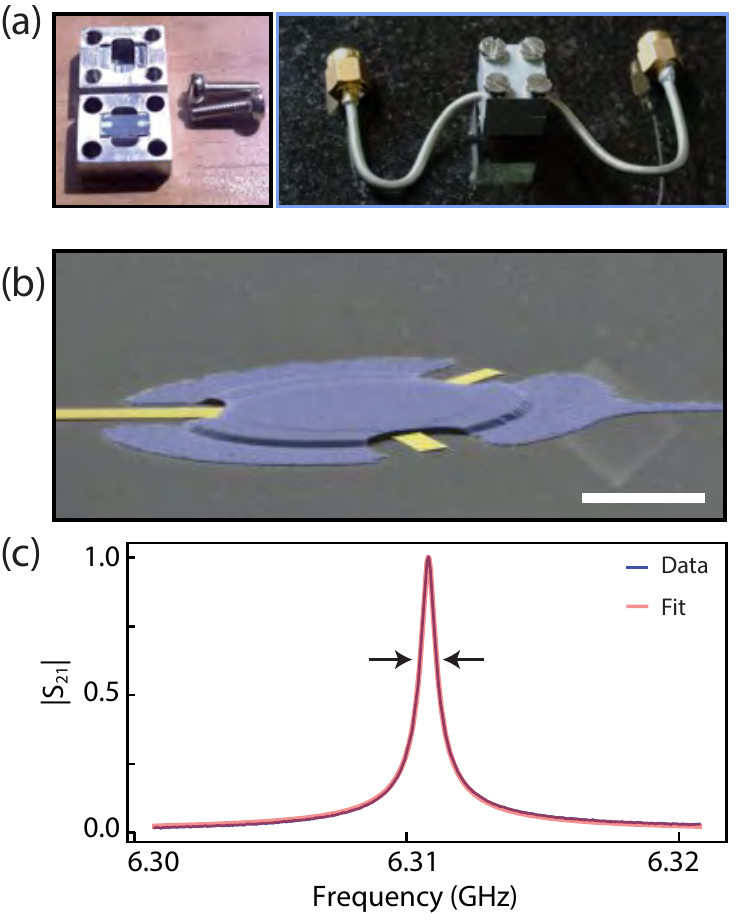}
	\caption{\textbf{Device}: (a) An image of a transmission cavity used to capacitively couple a mechanical resonator to the electromagnetic mode. Two M3 screws are shown to highlight the relative size. The right subpanel shows the complete assembly with SMA connectors. (b) False-color scanning electron microscope image of one of the drumhead resonator fabricated with aluminum on intrinsic Si substrate. The scale bar corresponds to 10~$\mu$m. The drumhead has a diameter of 22~$\mu$m, and is suspended $\approx 300$~nm above the bottom plate. (c) Measurement of transmission parameter of the optomechanical cavity at 20~mK (blue curve), along with the fitted curve (red) yielding a total dissipation rate $\kappa = 2 \pi \times $481~kHz as indicated by the arrows.}
	\label{fig-2}
\end{figure}

For experimental realization, rectangular 
waveguide cavities of size 7~mm $\times$ 4~mm $\times$ 7~mm were 
machined using 6061-T5 aluminum alloy, having diamond polished 
interior surfaces as shown in Fig.~\ref{fig-2}(a). A more detailed image of the cavity assembly is provided in SM.
The bare cavity has a fundamental mode frequency of 30.3~GHz, approximately. 
In order to lower the electromagnetic mode frequency to within 
our measurement bandwidth of 4-8 GHz, the mechanical resonator, 
serving the role of an added capacitor must be designed appropriately. 
To sufficiently load the electromagnetic mode capacitively, 
two drumhead-shaped mechanical resonators having their capacitances in parallel were designed.
Each resonator has a diameter of 22~$\mu$m, a gap of 300~nm from the bottom capacitor plate, and is patterned 
with a 7~$\mu$m wide connecting electrode. This accounts for approximately 
23~fF mechanically compliant capacitance, potentially lowering the cavity 
mode frequency to $\sim$~9.4 GHz. 

The mechanical resonator 
device was fabricated on a high resistivity (~$>$~10~k$\Omega$-cm) silicon substrate using 
multiple steps of optical lithography. The bottom and top electrodes 
were patterned on 100 nm aluminum films 
deposited by e-beam evaporation and dc sputtering methods, respectively. 
The deposition parameters for the top electrode were systematically 
investigated and optimized to obtain films of high tensile stress. 
A tilted angle image of one of the mechanical resonators in the 
drumhead shape is shown in Fig.~\ref{fig-2}(b). To increase the mechanical capacitance, we pattern two drumhead shaped mechanical resonators having their capacitance in parallel. Such a sample is then placed inside a two port 
microwave cavity. Patterned electrodes on the chip were wire-bonded to 
the cavity surface using aluminum wires of 25~$\mu$m diameter. 
A complete assembly of the device is shown in Fig.~\ref{fig-2}(a).

\section{Measurements and Analysis}\label{section:Measurements}

\begin{figure*}
\includegraphics[width = 130 mm]{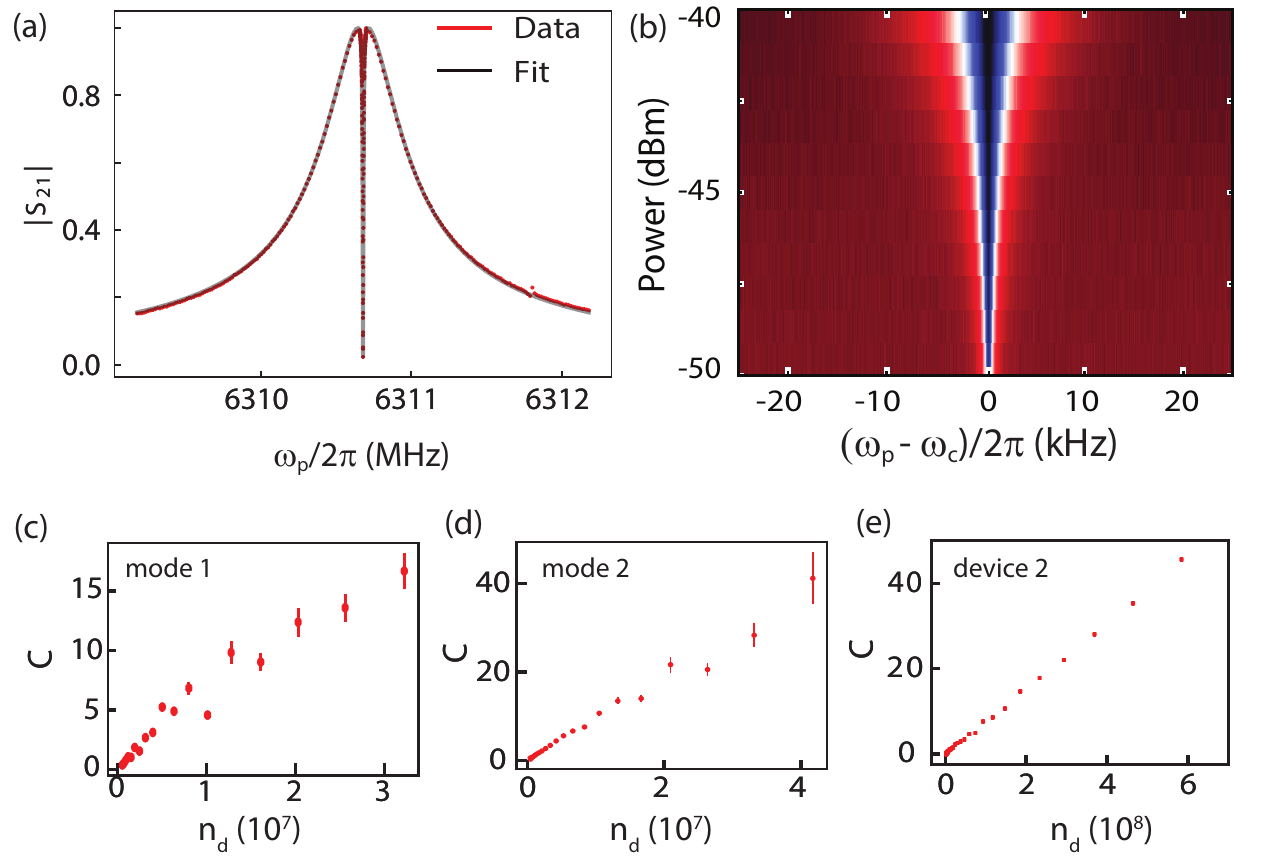}
\caption{(Color online) (a) Cavity transmission in the presence of a strong, red detuned pump signal at $\omega_{c}-\omega_{m}$ and a weak probe signal at $\omega_{c}$. (b) The colormap indicates variation in linewidth of the absorption feature at $\omega_c$ with drive (pump) power. (c), (d) and (e) are plots of cooperativity as a function of injected pump photons into the cavity. (c) and (d) were measured on device 1 that consisted of two mechanical resonators coupled to a 7 mm $\times$ 4 mm $\times$ 7 mm cavity. (e) was obtained for a single mechanical resonator (device 2) placed inside a 28 mm $\times$ 6 mm $\times$ 28 mm cavity. The highest values of cooperativity measured on the two devices differ due to the dynamic range of the two cavities. }\label{fig-3}
\end{figure*}

The assembled cavity was mounted to the mixing chamber plate of a dilution refrigerator and cooled down to the base temperature of 20~mK. A sufficiently attenuated input signal was used to drive the cavity and the transmitted signal was first amplified using a low-noise-amplifier and then measured using a vector network analyzer (VNA). In presence of the drumhead-shaped mechanical resonators, the fundamental mode of the cavity was recorded at $\omega_{c} \sim 2\pi \times 6.31$ GHz with a total line width of $\kappa = 2\pi \times 481$~kHz, as shown in Fig.~\ref{fig-2}(c). From simulations of the coupling ports, we estimated input and output coupling rates of 96~kHz and 330~kHz, respectively. Since most of the electric field remains in vacuum between the parallel plates of the capacitor, an internal cavity decay rate of 55~kHz is encouraging for future experiments in the quantum regime of motion. As the measured frequency is far off from the simulated value, we suspect the wirebonds ($\sim$ 2.5 mm long on each side) add a significant inductance to the cavity mode. Kinetic inductance, and reduction in the gap of the mechanical capacitor due to thermal contraction could also be playing a minor role.

The mechanical drumhead resonators and the optomechanical coupling were characterized by performing measurements in the optomechanically induced transparency (OMIT) configuration, wherein two microwave signals are used : a strong pump signal near $\omega_{c} - \omega_{m}$ and a weak probe signal near $\omega_{c}$, where $\omega_{c}$ and $\omega_{m}$ are the cavity and mechanical resonant frequencies, respectively \cite{weis_optomechanically_2010}. In the sideband resolved limit ($\omega_{m} > \kappa$), the presence of two signals detuned by $\omega_m$ exerts a beating radiation pressure force on the mechanical resonator, thereby driving it at its resonant frequency. The coherent motion of the mechanical resonator up-converts the sideband signal to  exactly the probe frequency $\omega_p$. The resulting optomechanical interaction between the microwave field and drum motion manifests in the form of a narrow absorption (OMIA) or a transmission (OMIT) window in an otherwise smooth cavity response.

Fig.~\ref{fig-3}(a) shows the cavity response measured in an OMIT 
configuration wherein, a sharp dip at the cavity frequency $\omega_{c}$ 
is observed. A small feature near  6.312~GHz corresponds to the 
second mechanical mode. In presence of a pump tone at the lower 
sideband ($\omega_{d} = \omega_c-\omega_m$), the cavity transmission 
is given by,
\begin{equation}
	S_{21}(\omega_p) = \frac{\sqrt{\eta_{L} \eta_{R}}}{\left(\frac{-i\Delta}{\kappa/2} + 1\right) + \left(\frac{C}{\frac{-i\Delta}{\gamma_{m} /2}+1}\right)} 
\end{equation}
where $\eta_{L}$ and $\eta_{R}$ are the coupling efficiencies 
of the input and output ports, $\Delta = \omega_{p} - \omega_{c}$, $C = \frac{4g_0^2n_d}{\kappa\gamma_m}$ is the optomechanical cooperativity, $n_d$ is the number of pump photons, $\kappa$ and $\gamma_{m}$ are 
the cavity and mechanical linewidths, respectively. Measurement results 
shown in Fig.~\ref{fig-3}(a) were fitted using Eq.~1. 
From these measurements, we determine the following parameters for the two mechanical modes: $\omega_{m1} = 2\pi \times 5.23$~MHz, $\omega_{m2} = 2\pi \times 7.35$~MHz, intrinsic linewidths of $\gamma_{m1} = 2\pi\times 250$~Hz and $\gamma_{m2} = 2\pi\times 200$~Hz.

At low pump powers, the linewidth of the OMIA feature is predominantly 
determined by intrinsic losses in the mechanical resonator, and the  
magnitude of the cavity transmission at the OMIA feature drops by a 
factor of $\frac{1}{1+C}$. At higher pump powers, the optomechanical 
damping dominates, leading to a broadening of the OMIA feature as the 
strength of pump tone is increased, as shown in Fig.~\ref{fig-3}(b). 
Thus, a measurement of the optomechanically-induced absorption feature 
can directly be used to estimate the optomechanical cooperativity. 
Figs.~\ref{fig-3}(c) and (d) show the variation of cooperativity
for the two mechanical modes. The number of pump photons $n_d$ were estimated 
from the known losses in the input chain. The errors bars result from
the numerical fits performed using Eq.~1 to the complete cavity response.
From the linear variation of $C$ with $n_d$ at low powers, we estimated a 
single photon coupling strength $g_{0}/2\pi$ of 8.0~Hz, and 6.5~Hz for the two mechanical modes, which are close to the simulated values of $\sim$~8.9~Hz, and $\sim$~7.5~Hz.
At high pump powers, the sub-linear variation of the cooperativity stems from 
the increase in cavity dissipation rate.

It is interesting to contrast this behavior with a device fabricated and assembled differently. A mechanical resonator of 17~$\mu$m diameter with a gap of $\sim$300~nm, patterned with a 50~$\mu$m wide connecting electrode was coupled to a 28 mm $\times$ 6 mm $\times$ 28 mm cavity. In this device, we use indium bump pads (instead of wire-bonds) 
to make galvanic connection to the cavity walls, which were tightly pressed by the other half.
A variation of cooperativity with $n_d$ for this device is shown in Fig.~\ref{fig-3}(e). With this approach, we are able to achieve a higher dynamic range, and even at the highest pump powers no sublinear variation in cooperativity was noticed. This observation suggests the potential role of current crowding near the cavity walls in wire-bonded devices.

\section{Tunable optomechanical device}\label{section:dc-bias}

\begin{figure*}
\includegraphics[width = 160 mm]{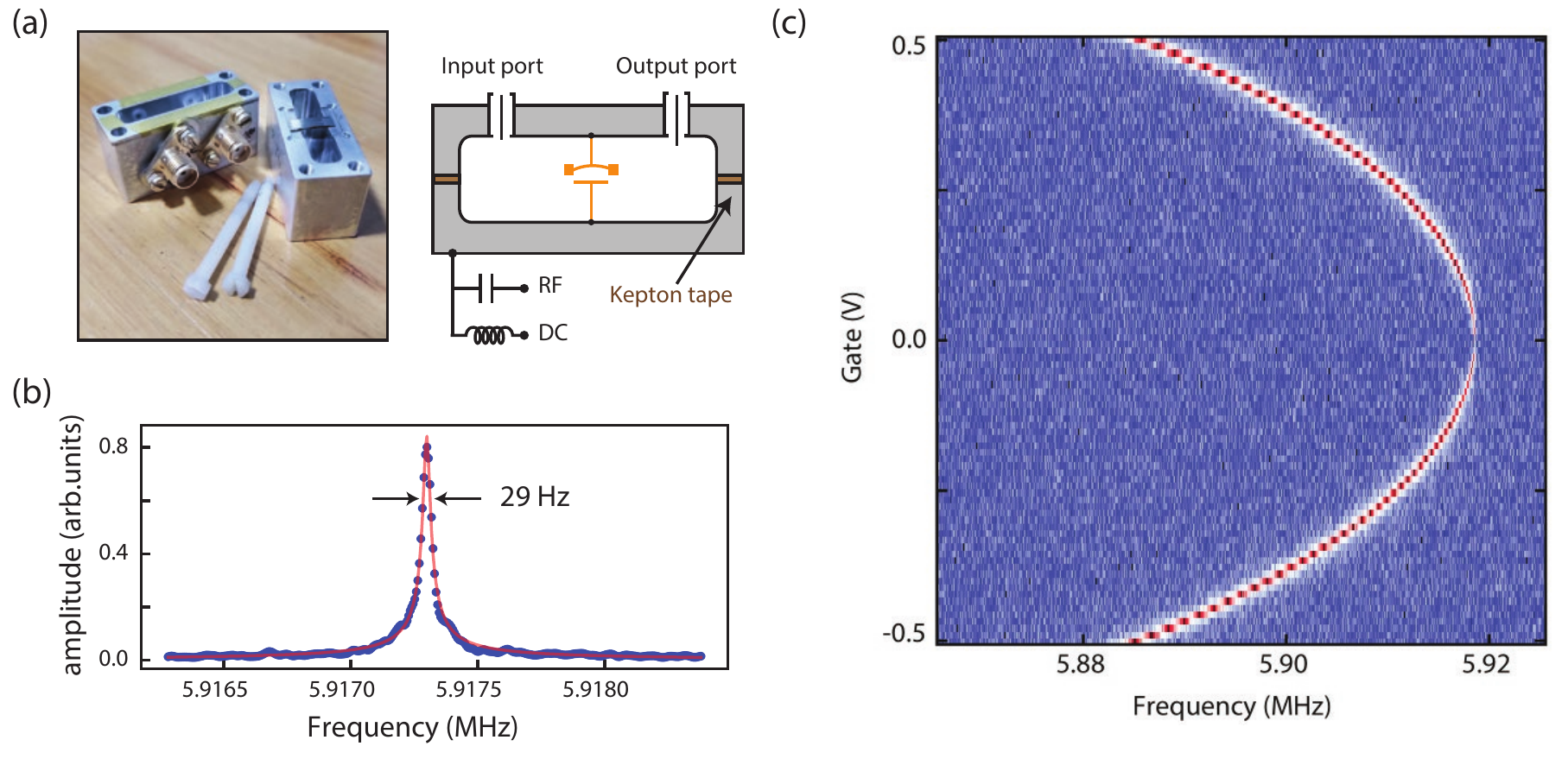}
\caption{(Color online) (a) A representative image showing the assembly of a DC-enabled optomechanical cavity, along with a schematic diagram. The two halves of the cavity are electrically isolated by a thin insulating spacer (thickness~$\sim$~10~$\mu$m). A bias-tee allows to add a low-frequency RF signal and a DC voltage to be applied across the mechanical resonator. (b) Response of the mechanical oscillator while directly driving it with a low-frequency RF signal. (c) Color plot of the demodulated signal with DC gate voltage and RF-drive frequency, showing the tunability of the resonant frequency of the mechanical resonator.}\label{fig-4}
\end{figure*}

The applicability of an optomechanical system can be extended 
by tuning the mechanically compliant capacitor with a DC voltage. 
Such systems could enable in-situ tuning of the mechanical modes and 
potentially have applications in deterministic capture of quantum information~\cite{andrews_quantum-enabled_2015}. Incorporating a DC bias in 
a microwave cavity while maintaining low internal losses at microwave 
frequencies is challenging. For broadband operations, it requires heavy 
on-chip reflective filtering on the DC-port \cite{andrews_quantum-enabled_2015}. 
For near-resonant operation, specially engineered devices exploiting the 
voltage nodes of the resonator have been attempted \cite{chen_introduction_2011, bosman_broadband_2015}.

The large capacitance between the two halves of a 3D waveguide cavity naturally offers a low-impedance environment around the electromagnetic mode of the cavity. Based on the physical dimensions of cavity-halves, we estimated a capacitance of 250~pF between them, suggesting a possibility of applying a DC-voltage via the two halves.
We introduce a DC voltage in our optomechanical device by electrically isolating the two halves of the cavity by a thin-insulating layer~\cite{cohen_split-cavity_2017}.
A representative image of cavity assembly, along with a schematic of such a device is shown in Fig.~\ref{fig-4}(a). Electrical isolation obtained between the two parts of the cavity enables the application of a DC voltage across the mechanical capacitor. Further, a low frequency RF signal can be added using a bias tee at low temperatures, 
facilitating a direct drive of the mechanical resonator given by the force $C^{'}_{m} V_{dc}V_{ac}$, where $V_{ac}$ and $V_{dc}$ are the amplitudes of RF and DC signals, respectively. 
For mechanical readout, a microwave tone at $\omega_c$ is injected into the cavity, 
and the transmitted signal is then demodulated and recorded at room temperature, 
while the mechanical resonator is driven by a low frequency RF signal. 
To reduce any RF-leakage to the output chain, we added two 4-8 GHz bandpass filters before the low noise amplifiers, yielding a 80~dB suppression of low frequency signals. We measured an insertion loss of $\sim$~34dB between the input microwave port and DC port of the cavity.

On characterization of microwave cavity, the internal linewidth was measured to be $\kappa_{i} \sim 1$ MHz. Further details on the cooperativity and single photon coupling strength of this system are provided in the SM. Fig.~\ref{fig-4}(b) 
shows the measurements from a device assembled to have a DC-voltage across the mechanical resonator. While directly driving it with a low-frequency RF signal, the mechanical response shows a linewidth of $\gamma_m/2\pi$~=~29~Hz.  The DC voltage allows us to tune the mechanical resonant frequency by 
electrostatic-pulling. Fig.~\ref{fig-4}(c) shows the parabolic variation 
in drum frequency with an applied gate voltage. Such a ``negative-dispersion'' of resonant frequency with DC gate voltage is well-understood due to capacitive-softening of the effective spring constant of the mechanical resonator \cite{kozinsky_tuning_2006}.

\section{Conclusions and Outlook} \label{section:conclusions}
	
To summarize, we described a scheme to couple nano/micro-mechanical resonators to a waveguide cavity. The design guidelines discussed here, help to minimize the parasitic capacitance, an important criterion to improve single-photon coupling strength. Our design, along with 
the ability to add a DC voltage can easily be extended to optomechanical systems 
with novel materials such as graphene, NbSe$_2$, BSCCO, etc to study their intrinsic performance, phase-transitions, or to further improve the 
single-photon coupling rate, while much simplifying the related nano-fabrication. Looking 
forward, by reducing the separation between the capacitor plates, we will be able to 
improve the single-photon coupling rate to $\approx$~200~Hz for a gap 
of 60 nm. Together with the large dynamic range of 3D cavities, this would let us perform experiments near the ultra-strong coupling limit, enabling the study of quantum behavior of massive mechanical resonators.

\section{Acknowledgments}

The authors would like to thank Mandar~Deshmukh for providing valuable inputs. 
This work was supported by the Department of Atomic Energy, under the Young Scientist Research Award. 
B.G. acknowledges support by the UGC under the D.S.~Kothari Fellowship program.
The authors acknowledge device fabrication facilities at CeNSE, IISc Bangalore, and central facilities at the Department of Physics funded by DST.

\end{document}